\pdfoutput=1
\documentclass[10pt,amsmath,amssymb,nofootinbib,twoside,twocolumn,
floats,floatfix,longbibliography,aps,prl]{revtex4-2}

\usepackage[british]{babel}
\usepackage{txfonts,mathrsfs,tensor}
\usepackage{natbib,textcase}
\usepackage{amsfonts,mathtools}

\usepackage{etoolbox}
\apptocmd{\sloppy}{\hbadness 10000\relax}{}{}
\usepackage{soul}
\usepackage[pdftex,pdfusetitle]{hyperref}
\hypersetup{colorlinks=true,linkcolor=red,citecolor=blue,filecolor=black,urlcolor=black,
pdfauthor={Edgardo Franzin, Stefano Liberati and Jacopo Mazza}}
\usepackage[capitalize]{cleveref}
\crefname{subequation}{Eqs.}{Eqs.}
\Crefname{subequation}{Equations}{Equations}
\labelcrefformat{subequation}{#2(#1)#3}

\usepackage{physics}

\makeatletter\g@addto@macro\bfseries{\boldmath}\makeatother%

\makeatletter
\renewcommand{\paragraph}{%
  \@startsection{paragraph}{4}%
  {\z@}{1ex}{-1em}%
  {\normalfont\normalsize}{}%
}%
\renewcommand{\section}[1]{\paragraph{\itshape#1.---}}
\makeatother

\def\be#1\ee{\begin{align}#1\end{align}}
\def\bse#1\ese{\begin{subequations}#1\end{subequations}}

\allowdisplaybreaks

\newcommand{\ie}{i.e.}
\newcommand{\eg}{e.g.}

\newcommand{\0}{\nonumber}

\renewcommand{\geq}{\geqslant}

\renewcommand{\leq}{\leqslant}

\newcommand{\uh}{\textsc{uh}}
\newcommand{\cu}{\textsc{quh}}

\usepackage{comment}

\begin{document}

\title{
Kerr Black Hole in Einstein--\AE{}ther Gravity}


\newcommand{\SISSA}{\affiliation{SISSA, International School for Advanced Studies, via Bonomea 265, 34136 Trieste, Italy}}
\newcommand{\InfnTS}{\affiliation{INFN, Sezione di Trieste, via Valerio 2, 34127 Trieste, Italy}}
\newcommand{\IFPU}{\affiliation{IFPU, Institute for Fundamental Physics of the Universe, via Beirut 2, 34014 Trieste, Italy}}

\author{Edgardo Franzin}
\author{Stefano Liberati}
\author{Jacopo Mazza}

\SISSA\IFPU\InfnTS

\begin{abstract}
While non-rotating black-hole solutions are well known in Einstein--\ae{}ther gravity, no axisymmetric solutions endowed with Killing horizons have been so far found outside of the slowly rotating limit. Here we show that the Kerr spacetime is also an exact vacuum solution of Einstein--\ae{}ther gravity in a phenomenologically viable corner of the parameter space; the corresponding \ae{}ther flow is characterised by a vanishing expansion. Such solution displays all the characteristic features of the Kerr metric (inner and outer horizons, ergoregion, etc.) with the remarkable exception of the causality-violating region in proximity of the ring singularity. However, due to the associated \ae{}ther flow, it is endowed with a special surface, inside the Killing horizon, which exhibits many features normally related to the universal horizon of the non-rotating solutions --- to which it tends in the limit of zero angular momentum.
Hence, these Kerr black holes are very good mimickers of their general relativistic counterparts while sporting important differences and specific structures. 
As such, they appear particularly well-suited candidates for future phenomenological studies.
\end{abstract}

\maketitle%

\section{Introduction}%
The current and future ability to probe gravity in strong, non-linear and dynamical regimes will allow to constrain several alternative scenarios, including some in which Lorentz invariance is broken.
Though usually considered a cornerstone of modern theoretical physics, reasons to enquire about Lorentz invariance abound: 
First of all, several approaches to quantum gravity seem to point to the possibility that it is either broken or deformed at high energies, only to be recovered at low energies as an accidental symmetry; 
moreover, renouncing Lorentz invariance makes it possible to construct quantum theories of gravity that are power-counting renormalisable and free of ghosts --- the prime example being Ho\v{r}ava gravity~\cite{horava_quantum_2009,blas_consistent_2010,blas_models_2011}.
Finally, while Lorentz invariance is well-tested in the matter sector, its violations in the gravitational sector are less constrained~\cite{liberati_tests_2013}.

Introducing Lorentz-breaking operators in a gravitational Lagrangian requires the identification of a preferred frame, which must be dynamical if background independence is to be preserved.
Such frame is typically embodied by a vector field $u_\mu$, usually dubbed \emph{\ae{}ther}, that is constrained to be everywhere timelike and of unit norm.
The \ae{}ther can therefore never vanish, in violation of Lorentz boost symmetry.

The most general Lagrangian preserving general covariance and, following an effective-field-theory reasoning~\cite{Withers:2009qg}, containing up to two derivatives defines the so-called Einstein--\ae{}ther theory~\cite{Jacobson:2000xp, Jacobson:2007veq}, whose Lagrangian reads
\be
\mathcal{L} = R + \mathcal{L}_\text{\ae} + \zeta \left(g_{\mu\nu}u^\mu u^\nu + 1\right),
\ee
where $\zeta$ is a Lagrange multiplier introduced to implement the unit-norm constraint.
The \ae{}ther's Lagrangian can be parametrised in different ways: here we shall adopt its ``hydrodynamical'' formulation introduced in Ref.~\cite{jacobson_undoing_2014}, though we use the opposite signature for the metric $(-,+,+,+)$: 
\be\label{KerrAE:eq:Lae}
\mathcal{L}_\text{\ae} = - \frac{1}{3}\,c_\vartheta\,\vartheta^2 - c_\sigma\,\sigma^2 - c_\omega\,\omega^2 + c_\mathfrak{a}\,\mathfrak{a}^2\,,
\ee
where $\vartheta^2,\ \sigma^2,\ \omega^2,\ \mathfrak{a}^2$ are the \ae{}ther's squared expansion, shear, twist and acceleration respectively.

Note that, when the twist vanishes, Frobenius' theorem ensures that the \ae{}ther is hypersurface orthogonal; in this case, Einstein--\ae{}ther theory reduces to (a subset of) the low energy limit of non-projectable Ho\v{r}ava gravity --- also known as khronometric theory~\cite{blas_consistent_2010,blas_models_2011}.

The couplings $c_\vartheta,\ c_\sigma,\ c_\omega$ and $c_\mathfrak{a}$ are tightly constrained by observations --- see \eg\ Refs.~\cite{oost_constraints_2018,gupta_new_2021,adam_rotating_2021} and references therein.
In particular, the multi-messenger observation of a binary neutron star merger~\cite{lv_gw170817,lvetal_multi-messenger_2017} allowed to constrain the difference between the speed of gravitational waves and that of light to within $10^{-15}$; this translates directly into a bound $c_\sigma \lesssim 10^{-15}$.
To date, only two regions of the parameter space have not been ruled out.
In the first, $c_\omega$ is unconstrained and possibly large, while $c_\vartheta$ and $c_\mathfrak{a}$ are small and equal up to even smaller corrections.
In the second region, \emph{both} $c_\omega$ and $c_\vartheta$ can be somewhat large, while $c_\mathfrak{a}$ is very small.
In either case, $c_\sigma$ is constrained to be so small that it is usually set to zero in computations.

\section{Black holes in Einstein--\ae{}ther gravity}%
Black holes are the standard laboratories to test modified theories of gravity and their strong-field regimes. Given that theoretical expectations and astrophysical observations strongly indicate that astrophysical black holes are endowed with angular momentum, significant attention has been devoted to rotating black hole solutions in extensions of general relativity.

Einstein--\ae{}ther theory admits black-hole solutions~\cite{eling_black_2006}, although their nature is quite subtle due to the Lorentz-violating character of the theory.
Indeed, Einstein--\ae{}ther encodes additional degrees of freedom with respect to general relativity, and since these modes propagate at different speeds, black holes generically exhibit multiple nested horizons.
Moreover, when including matter, the breaking of Lorentz invariance opens the way to the existence of modified, non-linear dispersion relations, which entail that high energy modes can propagate at arbitrarily high speed.  
Hence, in this context, metric horizons typically trap modes of low energy only.

However, in some situations black holes have been shown to exist at all energies due to the appearance of a novel causal surface delimiting a region from which not even infinite speed signals can escape, the so-called \emph{universal horizon}~\cite{barausse_black_2011,Blas:2011ni}.
To date, such universal horizons have been properly characterised only for globally foliated manifolds~\cite{bhattacharyya_causality_2016}, \ie\ in the case of a hypersurface-orthogonal \ae{}ther, and under the assumption of stationarity.\footnote{Although a local characterisation of universal horizons, suitable also for dynamical settings, can be provided~\cite{carballo-rubio_geodesically_2022}.}
In these cases, they represent special leaves of the preferred foliation characterised by the fact of being compact constant-radius surfaces, at which the \ae{}ther is orthogonal to the Killing vector of stationarity. 
Unfortunately, it is still unclear whether these structures can be extended to the more general case in which hypersurface orthogonality is absent.

Spherically symmetric black-hole solutions can be found analytically for special values of the couplings~\cite{berglund_mechanics_2012}; further numerical solutions were reported in Refs.~\cite{eling_black_2006,barausse_black_2011}. 
Notably, in a corner of the parameter space the solution consists in the Schwarzschild metric and an appropriate ``stealthy'' \ae{}ther flow~\cite{zhang_spherically_2020}.

Finding rotating solutions is more challenging. 
In the phenomenologically relevant sector, with small coupling constants, these solutions are expected to be parametrically ``close'' to the Kerr spacetime.
This expectation is realised in the slow rotation limit~\cite{barausse_slowly_2016}; and in the numerical rotating solutions found in Ref.~\cite{adam_rotating_2021}. 
Notably, these solutions exhibit metric horizons that are not Killing horizons.

To better understand the nature of rotating black holes in Einstein--\ae{}ther theory, we push this expectation to the extreme and look for solutions in which the metric is \emph{exactly} Kerr.
This strategy is motivated by the following remark: If one restricts the theory by setting 
\be
c_\omega= c_\sigma = c_\mathfrak{a} = 0 \, ,
\ee
\ie\ by switching off all couplings except for $c_\vartheta$, then any vacuum solution of general relativity is a solution of this theory too, provided the expansion of the \ae{}ther vanishes. 
Hence, in particular, the Kerr metric will be a solution if one can find an \ae{}ther of unit norm and such that $\nabla_\mu u^\mu = 0$ (and possibly some further constraints).
For the purpose of this paper, we will refer to this restricted version of Einstein--\ae{}ther theory as the ``minimal \ae-theory''.

In passing, we note that one could alternatively choose to set to zero all couplings except for $c_\sigma$ or $c_\omega$, then look for solutions of $\sigma_{\mu \nu} = 0$ or $\omega_{\mu \nu} = 0$, respectively; however, these choices would translate into a much more involved problem.
Similarly, one could choose to have $c_\mathfrak{a}$ as the only non-zero coupling and look for an \ae{}ther such that $\mathfrak{a}_\mu = 0$: the problem would be substantially simpler in this case, as it would reduce to that of finding timelike geodesics in the Kerr spacetime. 
Yet, the resulting \ae{}ther flow would be ``trivial'' and uninteresting for our purposes: for example, the projections of the \ae{}ther along the spacetime's Killing vectors would be constant and fixed by the boundary conditions. 
Hence, we resolve to focus on the minimal \ae{}-theory alone.

\section{The minimal \ae-theory\label{subsec:minimal}}%
Setting $c_\omega= c_\sigma = c_\mathfrak{a} = 0$ is a drastic reduction of arbitrariness. 
One might wonder, therefore, if the resulting theory is still relevant and viable.
We argue that the answer is in the positive, although further investigation on the matter might be required.

The full Einstein--\ae{}ther theory contains gravitational modes of three kinds: the familiar helicity-2 mode, corresponding to the usual gravitational waves; a helicity-1 mode; and a helicity-0 mode.
Each mode propagates {\em a priori} at a different speed ($c_T,\ c_V,\ c_S$ for the tensor, vector and scalar mode respectively) that can be computed studying linear perturbations around a given background and is set by the couplings of the theory.

In our minimal \ae{}-theory, the helicity-1 mode is non-dynamical. 
Indeed, since the \ae{}ther couples to gravity only through its expansion, the  helicity-1 mode never enters the action and thus it can be eliminated by a suitable gauge choice. 
Linearising around flat spacetime and a trivial \ae{}ther flow, one can then perform a computation along the lines of Ref.~\cite{jacobson_einstein-aether_2004} to find that $c^2_T = 1$, \ie\ the tensor mode moves at the speed of light; the fate of the helicity-0 mode is instead less clear, as this probably suffers from strong coupling.

Incidentally, the minimal \ae-theory is highly reminiscent of what the authors of Ref.~\cite{franchini_relation_2021} have called minimal khronometric theory --- the phenomenologically motivated restriction of khronometric theory.\footnote{The minimal khronometric theory is in turn related to a cuscuton theory~\cite{afshordi_cuscuton_2009}, cf.~Ref.~\cite{franchini_relation_2021}.}
The difference between the two minimal theories is that in the \ae-theory the twist coupling $c_\omega$ is set to zero, while in the khronometric theory the \ae{}ther is hypersurface orthogonal and its twist therefore vanishes by construction.

Such minimal khronometric theory has been shown to be indistinguishable from general relativity in all phenomenological applications so far considered.
In particular, in spherically symmetric stars and black holes the khronon is found to have a non-trivial profile and yet not to backreact on the geometry --- see the relevant discussion in Ref.~\cite{franchini_relation_2021} and references therein.

Hence, our preliminary conclusion is that the minimal \ae-theory is motivated and phenomenologically viable, though caution is advised.
For the purpose of this paper, however, the simplifications of the minimal theory are only needed insofar as they render the equations tractable analytically.
More generally, one could think of these solutions as the zeroth order of an expansion in the other coupling constants.
Thus, the pressing question becomes to understand if the solutions of the full theory \emph{can} in fact be expressed in such a Taylor-series fashion; or, alternatively, if the limit $c_{\omega,\ \sigma,\ \mathfrak{a}} \to 0$ is continuous in the space of the  solutions.

\section{The Kerr solution in the minimal \ae-theory}%
Let us then focus on the minimal \ae-theory and start by assuming that the metric is the Kerr one.
Adopting Boyer--Lindquist coordinates $\left(t,r,\theta,\phi \right)$, the Killing vectors associated to the spacetime's stationarity and axisymmetry are trivially
\be
\chi^\mu &= \left( 1,0,0,0 \right)^\mu \, , \\
\psi^\mu &= \left( 0,0,0,1 \right)^\mu \,.
\ee
and the Kerr metric reads
\be
ds^2 &= - \left( 1-\frac{2Mr}{\Sigma} \right) \dd{t^2} - \frac{4M r a \sin^2\theta}{\Sigma} \dd{t} \dd{\phi} \0 \\
&+ \frac{\Sigma}{\Delta} \dd{r^2} + \Sigma \dd{\theta^2} + \frac{A \sin^2\theta}{\Sigma} \dd{\phi^2} \, ,
\ee
where
\be
\Delta &= r^2 + a^2 - 2 M r \, , \quad \Sigma = r^2 + a^2 \cos^2\theta\, , \0 \\
A &= (r^2 + a^2)^2 -\Delta a^2 \sin^2 \theta\, , \quad a=J^2/M .
\ee

We wish to find an \ae{}ther flow that is Lie-dragged by the same Killing vectors and that has a vanishing expansion
\be\label{eq:0expansion}
\nabla_{\!\mu} u^\mu = 0\, ,
\ee
given that, as argued, this will result in a solution of the minimal \ae-theory.
To this aim, we start by assuming that the components of the \ae{}ther depend on $r$ and $\theta$ only, so that $u_\mu$ is Lie-dragged along the Killing vectors.
Then \cref{eq:0expansion} becomes
\be\label{eq:0expKerr}
\partial_r \left(\Delta u_r \right)
+ \frac{1}{\sin\theta}\,\partial_\theta \left( u_\theta \sin\theta \right) =0\,.
\ee

\Cref{eq:0expKerr} is under-determined.
One can approach it by separation of variables and then proceed to impose some boundary conditions, but this would still not pick a unique solution. 
One is therefore free to make some arbitrary choices, \eg\ set to zero the component $u_\theta$. 
Such choice is consistent with the spherically symmetric limit, as well as the requirement that, at infinity, the \ae{}ther becomes aligned with the timelike Killing vector; but is otherwise arbitrary and made here for simplicity only. Different choices will therefore lead, in general, to other solutions.

The ensuing solution is particularly simple and given by
\be
u_r = -\frac{M^2 \Theta(\theta) }{\Delta}\,,\quad u_\theta=0\,,
\ee
where $\Theta(\theta)$ is an arbitrary function.
The factor $M^2$ is purely conventional, while the sign is such that if $\Theta$ is taken to be positive then $u_r$ is negative at infinity. 
Note that, at infinity, $u_r \sim r^{-2}$. 
It is important to note that with this \ae{}ther it is impossible to satisfy the hypersurface-orthogonality constraint $u_{[\mu} \nabla_{\!\nu} u_{\rho]} =0$. 
This solution therefore exhibits a non-vanishing twist.

We can then solve the unit-norm constraint to express one of the two remaining components in terms of the other.
For simplicity, let $u_\phi = 0$, then
\be\label{eq:uchi2}
u_t^2 = \frac{\Sigma \Delta + M^4 \Theta^2}{A}\,  
\ee
and $u_t = \pm \sqrt{u_t^2}$: the minus sign ensures that the \ae{}ther is aligned with the timelike Killing vector at infinity, as is usually assumed; however, if $u_t$ has a zero somewhere, then it must change sign upon crossing it in order to ensure that the \ae{}ther be of class $\mathcal{C}^1$~\cite{del_porro_time_2022}.

The full solution thus reads
\be\label{eq:uDown}
u_\mu = \left( \mp \sqrt{\frac{\Sigma \Delta + M^4 \Theta^2}{A}}, - \frac{M^2 \Theta }{\Delta}, 0, 0\right)_\mu\, , 
\ee
or, raising the index,
\be\label{eq:uUp}
u^\mu = \left( - \frac{A}{\Sigma\Delta} u_t, - \frac{M^2 \Theta }{\Sigma}, 0, - \frac{2Mra}{\Sigma\Delta }u_t \right)^\mu \, .
\ee
Note that the choice $u_\phi = 0$ means that an observer comoving with the \ae{}ther has zero Killing angular momentum, $u_\mu \psi^\mu = 0$; this \ae{}ther is nonetheless rotating, in some sense, since $u^\phi \neq 0$.
In particular,
\be
\frac{u^\phi}{u^t} = \frac{2Mra}{A} = \Omega(r,\theta)\, ,
\ee
where $\Omega(r,\theta)$ is the angular velocity of frame dragging.

The component $u_r$ appears singular at the Killing horizons, where $\Delta =0$. 
As is well known, some components of the metric also appear ill-behaved at those points, but this singularity is merely a coordinate artefact and can be removed by changing coordinates.
A coordinate chart in which the \ae{}ther is manifestly regular at the outer Killing horizon is the one provided by ingoing Kerr coordinates~\cite{wald_general_1984}, in which the metric is regular at the future Killing horizon. 
In these coordinates, however, the \ae{}ther is still singular at the inner horizon; it becomes regular there when expressed in the closely related outgoing Kerr coordinates~\cite{wald_general_1984}, which also render the metric regular at the past Killing horizon.

A further check that the \ae{}ther flow is in fact regular consists in computing $\vartheta^2,\ \sigma^2,\ \omega^2$ and $\mathfrak{a}^2$.
These quantities completely characterise the flow and, being scalars, do not depend on the particular choice of coordinates.
Except for the expansion, which vanishes by construction, their explicit expressions are rather cumbersome and for this reason we omit them here; however, we have checked that they are finite at $\Delta = 0$.

These scalars do reveal the existence of a singularity, at which they diverge, located at $\Sigma = 0$.
This is exactly the location of Kerr's ring-like singularity.

\section{Fixing \texorpdfstring{$\Theta(\theta)$}{Theta(theta)}\label{sec:Theta}}%
The function $\Theta(\theta)$ is arbitrary.
Since it was introduced \emph{via} separation of variables, it must not depend on any of the coordinates except for $\theta$;
however, it might depend on the spin $a$.

As mentioned above, in the absence of rotation black-hole solutions are known analytically.
One might therefore restrict the admissible forms of $\Theta$ by demanding that \cref{eq:uDown} reduces to a known solution in the limit of vanishing spin.

The non-rotating solution in the corner of the coupling space corresponding to the minimal \ae-theory consists in the Schwarzschild metric and
\be\label{eq:sphu}
u^\mu = \left( \mp \frac{\sqrt{F(r)+ r_\text{\ae}^4/r^4}}{F(r)} , - \frac{r_\text{\ae}^2}{r^2},0,0 \right)^\mu \, .
\ee
Here $F(r) = 1- 2M / r$ and $r_\text{\ae}$ is a constant of integration.
Hence, demanding that our solution reduces to \cref{eq:sphu} amounts to imposing that
\be
\lim_{a \to 0} M^2 \Theta (\theta) = r_\text{\ae}^2\, .
\ee

In the non-rotating case, the constant $r_\text{\ae}$ needs fine-tuning if the solution is to describe a black hole.
In particular, one has to impose $r_\text{\ae} \geq \sqrt[4]{27}M/2$ to ensure that the \ae{}ther is real-valued.
Moreover, since in spherical symmetry the \ae{}ther is automatically hypersurface-orthogonal, the Schwarzschild metric with the \ae{}ther of \cref{eq:sphu} constitute a solution of (the minimal) khronometric theory, too.
One typically picks 
\be
r_\text{\ae} = \sqrt[4]{27} \frac{M}{2}\, ,
\ee
so that the solution displays a universal horizon in addition to a Killing horizon.

Coming back to the rotating case, we also need to worry that the \ae{}ther might not be real-valued, since the expression we found for $u_t^2$ is not manifestly non-negative.
In the $r>0$ region, we may ensure that $u_t^2 \geq 0$ simply by demanding that
\be
\min(M^4 \Theta^2) \geq - \min(\Sigma \Delta )\, .
\ee

Remarkably, this bound is satisfied by the trivial choice 
\be 
M^2 \Theta = r_\text{\ae}^2 \, ,
\ee
if $r_\text{ae}$ satisfies the corresponding bound for the spherically symmetric case, \ie\ $r_\text{\ae} \geq  \sqrt[4]{27} M/2$ --- where, as said, the equality implies the existence of a universal horizon in the zero-spin limit, which disappears as soon as the spin is restored.

This motivates us to investigate a different choice --- admittedly a more complicated but arguably more interesting one.
A graphical inspection of the function $\Sigma \Delta $ reveals a Mexican-hat-like shape: at any given $\theta$, the function has a minimum in the radial direction located at one of the roots of the cubic equation
\be\label{eq:min}
\partial_r \left(\Sigma \Delta  \right) = 0\, .
\ee
The root corresponding to the minimum, which we will call $r_\cu$, is a continuous function of $\theta$ (actually, of $a^2 \cos^2 \theta$). 
Henceforth, we will use the subscript ``\cu'' to indicate that a quantity is evaluated at $r=r_\cu(\theta)$ --- the meaning of the acronym will become clear in due time.

We thus propose to choose
\be\label{eq:Theta}
M^2 \Theta = \sqrt{- \Sigma_\cu \Delta_\cu} \, .
\ee
In this way, $(\Sigma \Delta + M^4 \Theta^2)_\cu=0$ and therefore $(u_t)_\cu = 0$. 
Incidentally, note that 
\be
[ \partial_\mu(\Sigma \Delta + M^4 \Theta) ]_\cu = 0 \qq{but}
[\partial_\mu (u_t )]_\cu \neq 0 \,.
\ee

\section{Absence of closed timelike curves}%
We will soon investigate the consequences of the choice \cref{eq:Theta} in great detail, but first we shall make an important side comment that holds for any choice of $\Theta$.

As is well known, the maximal analytical extension of the Kerr spacetime includes a region in which the Boyer--Lindquist radius $r$ is negative.
Such region can be reached by physical observers who fall into the black hole and cross $r=0$ away from the equatorial plane, thus avoiding the ring-shaped spacetime singularity.

Remarkably, there exists a region in the negative-$r$ patch in which the function $A$, which is positive for $r>0$, becomes negative.
This fact is widely regarded as physically problematic because it is synonymous to the existence of closed timelike curves.
Indeed, one can easily show that the function $A$ is proportional to the norm of the Killing vector $\psi$ that generates rotations around the axis of symmetry: when $A<0$, said vector becomes timelike; but its integral curves are closed by construction.  
Hence, in the Kerr spacetime, $A<0$ entails the existence of closed timelike curves.

Moreover, in the present context the function $A$ appears explicitly in the \ae{}ther solution --- specifically, it is the denominator of $u_t^2$ in \cref{eq:uchi2}.
Since for $r<0$ the numerator of \cref{eq:uchi2} is strictly positive, in the region in which $A<0$ we also have $u_t^2 < 0$.
In other words, in our solution, the \ae{}ther becomes purely imaginary in the same region that is associated with the existence of closed timelike curves.
This feature is unavoidable and, notably, it cannot be removed by a careful choice of $\Theta$. 

The boundary of such region, identified as the locus of points at which $A=0$, is a singularity for the \ae{}ther. 
One can convince oneself of this by recalling that $u_t = u_\mu\, \chi^\mu$, which is divergent at $A=0$, is a genuine scalar and that the Killing vector field $\chi^\mu$ is well-behaved everywhere --- except at the spacetime singularity, where its norm diverges. 

One might wonder whether it might still be possible to define a solution in the region $A<0$, so that observers decoupled from the \ae{}ther might access it.
In principle, this could be achieved by performing, in the region $A<0$, a Wick rotation in the $t$ and $\phi$ coordinates: the resulting metric would have Euclidean signature and two Killing vectors, both with positive norm. 
The \ae{}ther would then be real-valued also in the region $A<0$, and still with negative norm.
However, decoupled observers entering the region $A<0$ would experience a change in the signature of the metric, from Lorentzian to Euclidean.
While it is not obvious that such transition would be classically forbidden, a signature change is normally associated to a divergent particle creation in quantum field theory~\cite{White:2008xr}.  

In conclusion, the above discussion seems to strongly suggest that the \ae{}ther flow of the Kerr solution requires an excision of the causally challenging region $A<0$ normally associated with the existence of closed timelike curves.

This result, though somewhat serendipitous, agrees with the expectation that theories with a preferred time direction should not admit closed causal curves.

\section{Interpreting the ``\cu''~surface}%
In theories with a preferred foliation, such as Ho\v{r}ava gravity, under the assumption of stationarity, the condition
\be\label{eq:uchi}
u_\mu \, \chi^\mu = 0
\ee
locally characterises a universal horizon~\cite{bhattacharyya_causality_2016}.

As mentioned, Einstein--\ae{}ther theory is generically not endowed with a preferred foliation, but merely with a preferred threading, \ie\ a preferred time direction.
Indeed, the solution we are focusing on displays a non-vanishing twist and is therefore not hypersurface-orthogonal.

Still, it would be tempting to try to extend the notion of universal horizon to the full Einstein--\ae{}ther theory.
Whether the condition \eqref{eq:uchi} remains a meaningful characterisation of such horizons even in the absence of hypersurface orthogonality, however, is far from clear.

An argument in favour of \cref{eq:uchi} relies on studying the behaviour of two-dimensional expansions, in line with the reasoning of Ref.~\cite{carballo-rubio_geodesically_2022}.
Consider a unit spacelike vector $s_\mu$, orthogonal to $u_\mu$ and such that its integral curves are purely radial and ``outgoing'' at infinity.
Explicitly
\be\label{KerrAE:eq:sUp}
s_\mu &= \left( -u_r \sqrt{-\frac{g^{rr}}{g^{tt}}} , -u_t \sqrt{-\frac{g^{tt}}{g^{rr}}} , 0, 0 \right)_\mu \, .
\ee
The quantity
\be
\vartheta^{(s)} = \left( g^{\mu \nu} + u^\mu u^\nu - s^\mu s^\nu \right) \nabla_\mu s_\nu
\ee
measures the rate of change, along $s_\mu$, of the cross-sectional area of the two-surfaces that are orthogonal to both $u_\mu$ and $s_\mu$.
When $\vartheta^{(s)}<0$, these surfaces are ``trapped'', since they shrink as one moves ``outwards'' in the \ae{}ther frame.
Hence, $\vartheta^{(s)} = 0$ is an alternative local characterisation of the universal horizon which seems not to rely on hypersurface orthogonality, at least explicitly.

For any metric satisfying the circularity condition\footnote{Circularity is a condition on the Killing vectors of a given spacetime that translates into a symmetry of the metric components under the simultaneous transformations $t \mapsto -t$ and $\phi \mapsto - \phi$; see \eg\ Ref.~\cite[Ch.~3]{Heusler}.} (as Kerr does) and for any \ae{}ther flow such that $u_\vartheta = u_\phi = 0$ (us in our case), we have
\be
\vartheta^{(s)} = - (u_\mu\, \chi^\mu) \frac{\partial_r \sqrt{g_{\theta \theta} g_{\phi \phi}}}{ \sqrt{-g} }  \, ,
\ee
so the zeroes of $\vartheta^{(s)}$ coincide with those of $u_\mu\, \chi^\mu$.

In Boyer--Lindquist coordinates, $u_\mu\, \chi^\mu = u_t$, hence the choice for $\Theta$ of \cref{eq:Theta} entails  $(u_\mu\, \chi^\mu)_\cu = 0$.
This is in fact our motivation for the choice of \cref{eq:Theta} and it suggests that the surface $r=r_\cu(\theta)$ could play the role of a universal horizon.

This interpretation is strengthened by inspecting the slowly rotating limit. 
Indeed, in this case \cref{eq:min} gives
\be
r_\cu = \frac{3}{2}M + \order{a^2}
\ee
and
\be
u_\mu &= \left(\mp \sqrt{1-\frac{2M}{r} + \frac{27}{16}\frac{M^4}{r^4}} , - \frac{3\sqrt{3} M^2}{r(r-2M)}, 0, 0  \right)_\mu \0\\
&+ \order{a^2} \, ,
\ee
which coincides with the corresponding spherically symmetric analytical solution up to $\order{a^2}$.  

The interpretation is further strengthened by considering the highly spinning limit, \ie\ $a \to M$.
In this limit, the \ae{}ther's twist vanishes and $r_\cu(\theta)$ becomes a constant-Boyer--Lindquist-radius surface that coincides with the (degenerate) Killing horizons.
This behaviour is thus highly reminiscent of so-called degenerate universal horizons~\cite{franchini_black_2017}.

For generic values of the spin, a closed-form expression for $r_\cu(\theta)$ could be found but would not be particularly informative, and for this reason we omit it here.
We can however state that this surface will lie between the two Killing horizons, since a radial minimum of $\Sigma \Delta $ must be located where $\Delta<0$. 
More specifically, since \cref{eq:min} can be written as
\be\label{KerrAE:eq:uh}
\Delta_\cu = - \Sigma_\cu \frac{r_\cu - M }{r_\cu}\, ,
\ee
we deduce that $r_\cu \geq M$; moreover, yet another rewriting of \cref{eq:min} tells us that
\be
r_\cu = \frac{3}{2}M - \frac{a^2}{2r_\cu } \left( 1+ \cos^2\theta \frac{r_\cu - M}{r_\cu} \right) \leq \frac{3}{2}M \, .
\ee
To summarise, 
\be
M  = \lim_{a \to M} r_\cu \leq r_\cu \leq \lim_{a \to 0} r_\cu = \frac{3}{2}M \, .
\ee

\begin{figure}[t]
    \centering
    \includegraphics[width=.45\textwidth]{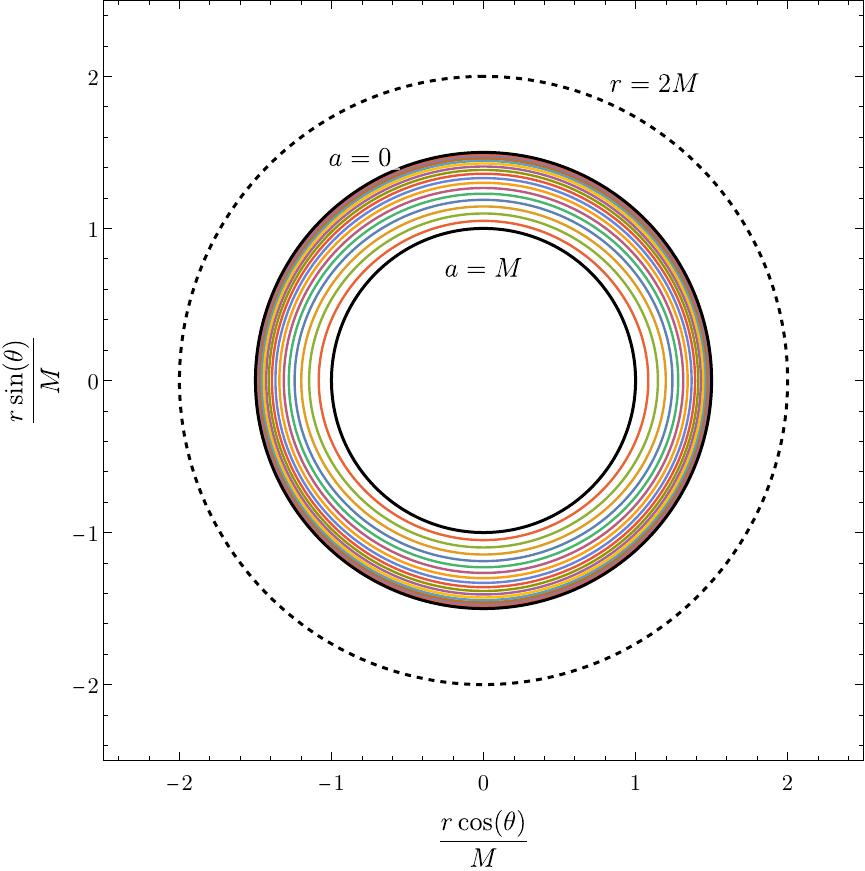}
    \caption[Plots \texorpdfstring{$r_\cu(\theta)$}{r_\cu(theta)}.]{Plots of $r_\cu(\theta)$ for several values of the spin $a$. The dotted line, which is reported for reference, corresponds to $r=2M$, \ie\ to the Killing horizon's radius of a Schwarzschild black hole of the same mass as our solution. Note that the curves look circles but, generically, they are not.}
    \label{fig:UH}
\end{figure}

Crucially, however, away from the particular cases $a=0$ and $a=M$ the dependence of $r_\cu(\theta)$ on the angle $\theta$ is concrete, albeit mild --- this will introduce considerable complications, as will become clear momentarily. 
A plot of the surfaces $r=r_\cu(\theta)$ for several values of the spin parameter is presented in \cref{fig:UH}.
Differentiating \cref{eq:min} with respect to $\theta$ gives
\be
\dv{r_\cu}{\theta} = - \dv{(a^2 \cos^2 \theta)}{\theta} \left[ 4r_\cu  - \frac{M \Delta_\cu}{(r_\cu - M)^2} \right]^{-1} \, ,
\ee
which vanishes at $\theta = 0,\, \pi/2,\, \pi$ but is non-zero otherwise.
(Note that the quantity in square brackets is positive).

Moreover, the surface $r=r_\cu(\theta)$ is \emph{not} orthogonal to the \ae{}ther. Its normal vector $n_\mu = \nabla_\mu(u_\nu\, \chi^\nu)$ can be written as 
\be\label{KerrAE:eq:n}
n_\mu &= \chi^\nu \left[ \nabla_{\mu} u_{\nu}-\nabla_{\nu} u_{\mu} \right] \nonumber \\
&= - \left( \mathfrak{a}_\nu \chi^\nu \right) u_\mu + \left( u_\nu \chi^\nu \right) \mathfrak{a}_\mu + 2 \omega_{\mu \nu} \chi^\nu \nonumber \\
&\overset{\cu}{=} - \left( \mathfrak{a}_\nu \chi^\nu \right) u_\mu + 2 \omega_{\mu \nu} \chi^\nu \, ,
\ee
\ie\ it has a component along the \ae{}ther and another component, orthogonal to the \ae{}ther, that is controlled by the twist.
We have checked that $n_\mu n^\mu < 0$, so $r=r_\cu(\theta)$ is a spacelike hypersurface generated by the two Killing vectors $\chi^\mu$ and $\psi^\mu$, and by $\rho^\mu = \left(0,\dv{r_\cu}{\theta}, 1, 0\right)^\mu$. 

Had the \ae{}ther been hypersurface-orthogonal ($\omega_{\mu \nu}=0$), the surface $r=r_\cu$ would have been a universal horizon and the projection $u_\mu \, n^\mu$ would have been interpreted as the horizon's surface gravity (up to a conventional normalisation factor).
In our case, instead, the misalignment between the \ae{}ther and the normal vector $n^\mu$, which is related to the lack of hypersurface orthogonality, strongly suggests that $r=r_\cu(\theta)$ \emph{cannot be interpreted as a true universal horizon}.

The reason for this is connected to the fact that the hypersurfaces generated by the integral curves of vectors that are orthogonal to the \ae{}ther are hypersurfaces of simultaneity with respect to the preferred time.
This is true by definition in khronometric theory, and also in Einstein--\ae{}ther theory --- the technical difference being that, in the absence of hypersurface orthogonality, these hypersurfaces do not constitute immersed submanifolds (cf.\ Ref.~\cite[App.~B.3]{wald_general_1984}).
Heuristically, we might regard these surfaces as generated by ``infinitely fast'' trajectories, \ie\ as ordinary causal trajectories in the limiting case in which their speed in the preferred frame becomes infinite.
The fact that the surface $r=r_\cu(\theta)$ is not one such hypersurface means that causal curves could cross it in both directions.  
However, let us look at this in more detail before drawing our conclusions.

\section{Infinite-speed trajectories}%
To investigate the trapping properties of the surface $r=r_\cu$, we may consider a generic curve $x^\mu(\lambda)$, with associated tangent vector $k^\mu=\dv*{x^\mu}{\lambda}$, and assume that the curve has ``infinite speed'' in the sense that $u_\mu k^\mu = 0$.
With an abuse of terminology, we will refer to this curve as an ``infinitely fast'' trajectory.

The vector $s_\mu$ of \cref{KerrAE:eq:sUp} can be complemented with
\be
\Big( e_{\hat{\theta}} \Big)^\mu &= \frac{1}{\sqrt{g_{\theta\theta}}} (0,0,1,0)^\mu \qq{and} \nonumber\\
\Big( e_{\hat{\phi}} \Big)^\mu &= \frac{1}{\sqrt{g_{\phi\phi}}} (0,0,0,1)^\mu
\ee
to obtain an orthonormal basis of the subspace of the tangent space that is orthogonal to $u_\mu$.
Together with $u_\mu$, these three vectors span the whole tangent space.
We may thus decompose
\be
k_\mu =  k_s s_\mu + k_{\Hat{\theta}}\Big( e_{\Hat{\theta}} \Big)_\mu+ k_{\Hat{\phi}} \Big( e_{\Hat{\phi}} \Big)_\mu\, ,
\ee
where $k_s,\ k_{\Hat{\theta}},\ k_{\Hat{\phi}}$ are given by the scalar product of $k^\mu$ with $s_\mu,\ \Big( e_{\hat{\theta}} \Big)_\mu$ and $\Big( e_{\hat{\phi}} \Big)_\mu$, respectively.

If $k^r \neq 0$, the map $r(\lambda)$ is invertible and $r$ can be used as a coordinate along the trajectory \emph{in lieu} of $\lambda$.
We may focus on the $(t,r)$-plane, where the most interesting motion happens, and neglect the angular motion.
The trajectory has tangent derivative
\be
\dv{t}{r} = \dv{t}{\lambda} \dv{\lambda}{r} = \frac{k^t}{k^r} \, .
\ee
Specifically, since $\Big( e_{\hat{\theta}} \Big)^t = \Big( e_{\hat{\phi}} \Big)^t = 0$,
\be
\frac{k^t}{k^r} &= \frac{k_s s^t}{k_s s^r} \nonumber \\
&= - \frac{u_r }{ u_t} \,;
\ee
in passing from the first to the second line, we have used the fact that
\be
u_\mu s^\mu = 0 \Rightarrow \frac{s^t}{s^r} = - \frac{u_r}{u_t}\, .
\ee

Hence, the trajectory has an asymptote at $u_t=0$, which is universal in the sense that it does not depend on $k_\mu$ as long as $k^r\neq 0$.
In particular, curves starting in the region $u_t>0$, \ie\ at $r < r_\cu$, can never cross $r=r_\cu(\theta)$ in the outward direction.

\begin{figure}[t]
    \centering
    \includegraphics[width = .45\textwidth]{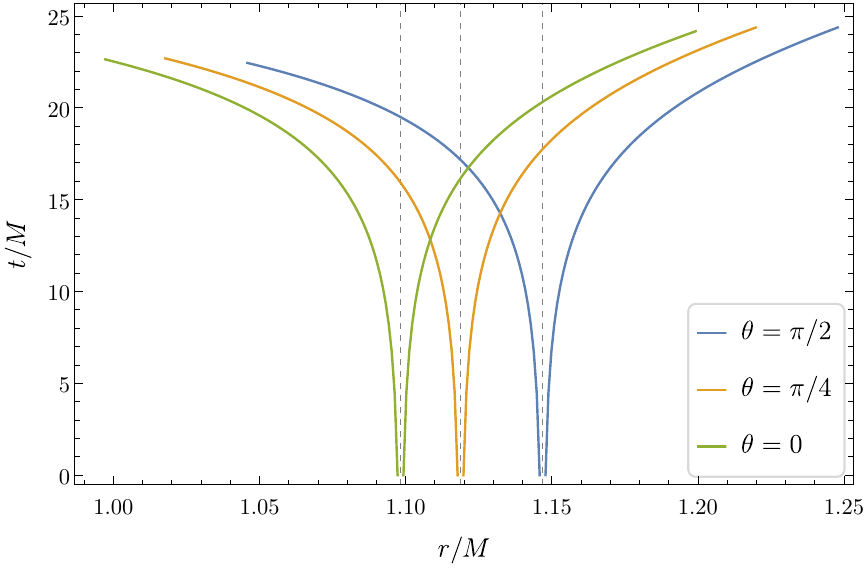}
    \caption[Peeling at \texorpdfstring{$r=r_\cu(\theta)$}{r=r_\cu (theta)}.]{Integral lines of the vector $s^\mu$, at different angles $\theta$. The vertical lines mark $r=r_\cu(\theta)$. The spin is $a=0.9M$.}
    \label{KerrAE:fig:intS}
\end{figure}

\Cref{KerrAE:fig:intS} displays the integral curves of $s^\mu$ in the $(t,r)$ plane in the vicinity of $r=r_\cu$ (since these coordinates are ill-behaved at the Killing horizons, a coordinate change is needed if one wishes to extend these plots beyond those surfaces).
The behaviour of such curves is highly reminiscent of the ``peeling'' typically witnessed when studying the motion of null rays close to a trapping horizon in general relativity, as well as the hypersurfaces of simultaneity in the vicinity of a universal horizon in khronometric theory.

The remaining case $k^r=0$, corresponds to ``particles'' moving only in the angular directions.
There exist curves that start at $r < r_\cu$ and, moving in the direction of $\Big( e_{\hat \theta} \Big)^\mu$, cross $r=r_\cu$ outwards.
By continuity, it seems possible that there also exist causal curves that similarly exit $r=r_\cu$. 

However, the mere existence of infinite-speed curves piercing $r=r_\cu$ does not necessarily mean that orbits of physical particles would do so too. 
To understand why this might not be the case, we need to characterise these ``purely angular'' trajectories.
First of all, let us note that
\be
k^r = k_s s^r\, .
\ee
Since $s^r\propto u_t$, this component can be zero at $r=r_\cu$ even if $k_s \neq 0$.
Leaving this particular case aside, for all $r\neq r_\cu$ we have that $k^r = 0 \Leftrightarrow k_s = 0$.

Because of the symmetries of the spacetime, there are two constants of motion, each associated to one of the Killing vectors.\footnote{This point is subtle. These trajectories are not geodesics, so it is not obvious that the Killing vectors generate conserved quantities. In the present case they do, in a sense that is made precise in Ref.~\cite{cropp_ray_2014}.}
The two quantities are $k_\mu \chi^\mu = k_t = -E$ and $k_\mu \psi^\mu = k_\phi = L$.
Note that $k_\phi= k_{\hat{\phi}} \sqrt{g_{\phi \phi}}$.

Consider in particular the conservation equation for $k_t=-E$:
\be
-E &= k_s s_t + k_{\hat{\phi}} \Big( e_{\hat{\phi}} \Big)_\mu \nonumber \\
&= k_s s_t - L \Omega\, ,
\ee
where $\Omega =- g_{t \phi}/g_{ \phi \phi}$ as above.
If we set $k_s = 0$ at some point, since $E$ and $L$ are constant while $\Omega$ is a function of $\theta$ (and $r$, which however we assume fixed), the only option is that $L=0$ and therefore also $E=0$.
But this entails that $k_s$ must vanish everywhere along the curve and therefore the curve can never reach infinity.
Moreover, if $k_s$ vanishes, then $k^t=k_s s^t=0$ everywhere too, \ie\ this curve never advances in the Killing time. 
Though in no way conclusive, this analysis seems hence to suggest that the only curves that pierce through $r=r_\cu$ do not correspond to (the infinite-speed limit of) physical trajectories.

Therefore, the upshot of the present discussion is that the status of the surface $r=r_\cu(\theta)$ is unclear.
It is not truly a universal horizon, because there seem to exist causal curves that cross it; yet, it exhibits some of its characteristic features, namely: it reduces to a universal horizon in both the $a\to 0$ and $a\to M$ limits, and it traps all infinite-speed signals whose tangent vectors have a non-vanishing radial component.  
For this reason, and for lack of a better terminology, we have called this surface ``\cu'' as in ``quasi universal horizon''.

\section{A note on the surface gravity}%
Universal horizons in globally foliated manifolds emit Hawking radiation in a way similar to horizons in general relativity~\cite{del_porro_time_2022,del_porro_gravitational_2022,del_porro_hawking_2023}.
For very high energy particles the temperature of such radiation is set by the universal horizon's surface gravity, defined as
\be
-2 \kappa_\text{h.o.} =  (\mathfrak{a}_\mu \, \chi^\mu)_\uh 
\ee
(the subscript ``h.o.''\ stands for ``hypersurface-orthogonal'', while the subscript ``UH'' means that the quantity is evaluated at the universal horizon).
Since, in that context, universal horizons are leaves of the preferred foliation, their surface gravity is necessarily constant along the horizon and a zeroth law of black hole mechanics automatically holds.

Clearly, it is not obvious that the surface gravity of a universal horizon with non-vanishing twist --- supposing it exists --- should be defined in the same way. 
Indeed, not surprisingly, the quantity $\kappa_\text{h.o.}$ computed for our solution is \emph{not} constant on the surface $r=r_\cu$.

The decomposition of the normal vector in \cref{KerrAE:eq:n} suggests an alternative definition
\be \label{KerrAE:eq:kappaN}
2 \kappa_n = \left[ \left( \frac{n^\mu}{\sqrt{-n^\alpha n_\alpha}}\right) \nabla_\mu (u_\nu \chi^\nu) \right]_\cu = \sqrt{-n^\nu n_\nu}_\cu 
\ee
(the subscript ``$n$'' in $\kappa_n$ now stands for ``normal''),
where now
\be
\kappa_n^2 = \kappa^2_\text{h.o.} - \kappa^2_\omega
\ee
with
\be
\kappa^2_\omega =  \left[\omega_{\mu \nu} \chi^\nu \omega^\mu_{\ \rho} \chi^\rho \right]_\cu \, .
\ee
In the slowly rotating limit, we find
\be
-2 \kappa_\text{h.o.} &= \frac{2}{3M}\sqrt{\frac{2}{3} } \left[1 + \frac{-17+\cos^2\theta}{27 M^2} a^2 + \order{a^4} \right] \\
-2 \kappa_n &= -2 \kappa_\text{h.o.} + \order{a^4}\, ,
\ee
while 
\be
\kappa_\omega = \order{a^2} \, .
\ee

\begin{figure}[t]
    \centering
    \includegraphics[width = .45\textwidth]{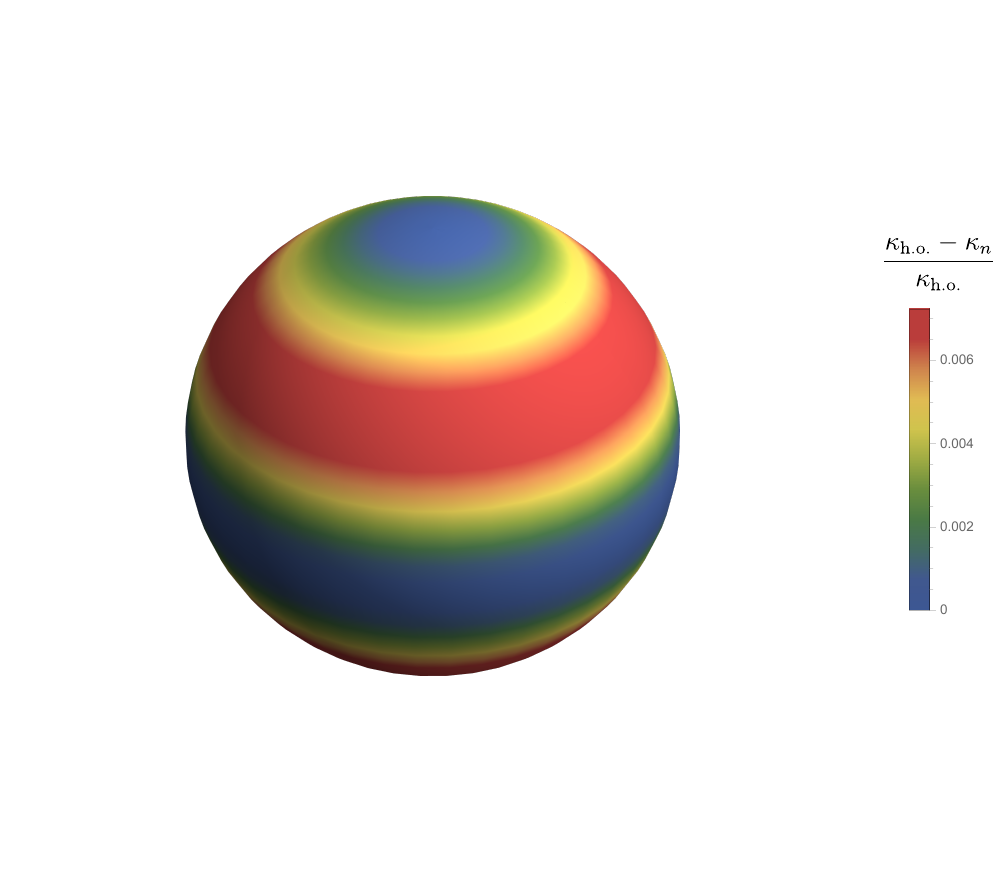}
    \caption[Relative difference in surface gravities.]{Relative difference of the two definitions of surface gravity, as a function of position on the surface $r=r_\cu(\theta)$. The spin is $a = 0.9M$.}
    \label{KerrAE:fig:Dk3D}
\end{figure}

The definition \eqref{KerrAE:eq:kappaN} is not constant along $r=r_\cu$ either, but it has the advantage of encoding information on the twist in an obvious way.
The two definitions coincide at the poles and at the equator, where the twist vanishes.
For arbitrary angles, their difference can be computed numerically: though growing with the spin, the relative difference is always $\lesssim 1 \%$.
An example for $a=0.9M$ is plotted in \cref{KerrAE:fig:Dk3D}. 

\section{Discussion}%
In this paper, we tackled the problem of finding rotating solutions in Einstein--\ae{}ther gravity by focusing on a restricted version of the theory, which we referred to as minimal \ae{}-theory.
We have assumed the metric to be that of Kerr and derived a compatible \ae{}ther flow. 
Though the equations of motion admit several solutions, we only analysed the simplest of them.

Our solution depends on a free function $\Theta$ of the polar angle~$\theta$.
Inspired by analogy with the known non-rotating solution, we proposed two ways in which such function can be fixed.
The first of such choices is very simple; the second is admittedly more involved but arguably more interesting.
A key feature of the second of such choices is the presence of a three-dimensional spacelike hypersurface that resembles in many ways a universal horizon, although it is not exactly one; we have named this hypersurface ``quasi universal horizon''.

The relevance of this last remark lies in the fact that, so far, universal horizons have only been described in the context of globally foliated manifolds, \ie\ they are associated to hypersurface-orthogonal \ae{}ther flows.
Our solution does not belong to this class, and indeed it does not display a full-fledged universal horizon; however, our work seems to hint to the possibility that some of the most salient features of universal horizons might be captured by the weaker notion of quasi universal horizons, which would only trap physical causal orbits but not all causal curves.

Since, in theories admitting violations of Lorentz invariance, universal horizons are the only true causal horizons, such questions are key in determining under what circumstances theories of this kind admit true black holes and respect the weak cosmic censorship conjecture. 

As a final comment, let us stress that in Ref.~\cite{del_porro_hawking_2023} it has been clarified that for large black holes the actual temperature observed at infinity is still dominated by the Killing horizon's surface gravity, with small deviations from exact thermality associated to the non-linear dispersion relation of matter. 
Hence, the weak dependence of the surface gravity on the polar angle might be physically irrelevant for the thermodynamic behaviour of black holes whose mass is larger than the Lorentz-breaking scale $\Lambda$~\cite{del_porro_hawking_2023}. 
Indeed, as long as the behaviour of the physical trajectories at the quasi universal horizon can be used to fix the vacuum state (without the need to impose boundary conditions at the singularity), these solutions might be thermodynamically viable for macroscopic black holes (and in any case for $M\sim\Lambda$ one might get corrections to the solution from the ultraviolet completion of the gravitational theory). 
This is again related to the possibility or impossibility of physical particles propagating from the singularity up to infinity. 
So, also for this reason, a more in-depth analysis of physical obits in this solution appears to be in order.

In conclusion, we have here provided a solution of a phenomenological viable corner of Einstein--\ae{}ther gravity. 
In such solution the metric is exactly Kerr while a non-trivial \ae{}ther flow leads to several interesting new features. 
We hence hope that this work will stimulate further theoretical exploration and provide a test bed for the phenomenology of rotating solutions in Einstein--\ae{}ther gravity.

\bigskip

\begin{acknowledgments}

The authors wish to thank Enrico Barausse, Francesco Del Porro, Mario Herrero-Valea, Ted Jacobson, and David Mattingly for their precious comments.
The authors acknowledge funding from the Italian Ministry of Education and Scientific Research (MIUR) under the grant PRIN MIUR 2017-MB8AEZ\@.
\end{acknowledgments}


\bibliography{refs}

\end{document}